\documentclass{aa}  
\usepackage{graphicx}
\usepackage{lscape}
\usepackage{longtable}
\usepackage{soul}
\usepackage{lipsum}
\usepackage{hyperref}

\usepackage{chngcntr}

\usepackage{txfonts}
\usepackage{float}

\newcommand{\Msun}{$M_{\odot}$}

\newcommand{\mic}{$\mu$m}

\usepackage{xcolor}

\begin{document}

   \title{Full L- and M-band high resolution spectroscopy of the S CrA binary disks with VLT-CRIRES+\thanks{Based on observations collected at the European Southern Observatory under ESO programme 107.22T7}}

   \author{Sierra L. Grant\inst{1}
          \and
          Giulio Bettoni\inst{1,2}
          \and
          Andrea Banzatti\inst{3}
          \and 
          Ewine~F. van~Dishoeck\inst{1,2}
          \and
          Sean Brittain\inst{4}
          \and
          Davide Fedele\inst{5}
          \and 
          Thomas Henning\inst{6}
          \and
          Carlo F. Manara\inst{7}
          \and
          Dmitry Semenov\inst{6,8}
          \and
          Emma Whelan\inst{9}
          }

   \institute{Max-Planck Institut f\"{u}r Extraterrestrische Physik (MPE), Giessenbachstr. 1, 85748, Garching, Germany
        \and
        Leiden Observatory, Leiden University, 2300 RA Leiden, the Netherlands
        \and
        Department of Physics, Texas State University, 749 N Comanche Street, San Marcos, TX 78666, USA
        \and 
        Clemson University, 118 Kinard Laboratory, Clemson, SC 29631, USA
        \and 
        INAF-Osservatorio Astrofisico di Arcetri, L.go E. Fermi 5, 50125 Firenze, Italy
        \and
        Max-Planck-Institut f\"{u}r Astronomie (MPIA), K\"{o}nigstuhl 17, 69117 Heidelberg, Germany
        \and
        European Southern Observatory, Karl-Schwarzschild-Strasse 2, 85748 Garching bei München, Germany
        \and
        Department of Chemistry, Ludwig-Maximilians-Universit\"{a}t, Butenandtstr. 5-13, D-81377 M\"{u}nchen, Germany
        \and
        Maynooth University Department of Experimental Physics, National University of Ireland Maynooth, Maynooth Co. Kildare, Ireland    
             }

  \abstract
  % context heading (optional)
  % {} leave it empty if necessary  
   {The Cryogenic IR echelle Spectrometer (CRIRES) instrument at the Very Large Telescope (VLT) was in operation from 2006 to 2014. Great strides in characterizing the inner regions of protoplanetary disks were made using CRIRES observations in the L and M band at this time. The upgraded instrument, CRIRES+, became available in 2021 and covers a larger wavelength range simultaneously.}
  % aims heading (mandatory)
   {Here we present new CRIRES+ Science Verification data of the binary system S Coronae Australis (S CrA). We aim to characterize the upgraded CRIRES+ instrument for disk studies and provide new insight into the gas in the inner disk of the S CrA N and S systems.}
  % methods heading (mandatory)
   {We analyze the CRIRES+ data taken in all available L- and M-band settings, providing spectral coverage from 2.9 to 5.5 $\mu$m.}
  % results heading (mandatory)
   {We detect emission from $^{12}$CO (v=1-0, v=2-1, and v=3-2), $^{13}$CO (v=1-0), hydrogen recombination lines, OH, and H$_2$O in the S CrA N disk. In the fainter S CrA S system, only the $^{12}$CO v=1-0 and the hydrogen recombination lines are detected. The $^{12}$CO v=1-0 emission in S CrA N and S shows two velocity components, a broad component coming from $\sim$0.1 au in S CrA N and $\sim$0.03 au in S CrA S and a narrow component coming from $\sim$3 au in S CrA N and $\sim$5 au in S CrA S. We fit local thermodynamic equilibrium slab models to the rotation diagrams of the two S CrA N velocity components and find that they have similar column densities ($\sim$8$\times$10$^{16}$-4$\times$10$^{17}$ cm$^{-2}$), but that the broad component is coming from a hotter and narrower region.}
  % conclusions heading (optional), leave it empty if necessary 
   {Two filter settings, M4211 and M4368, provide sufficient wavelength coverage for characterizing CO and H$_2$O at $\sim$5 $\mu$m, in particular covering low- and high-$J$ lines. CRIRES+ provides spectral coverage and resolution that are crucial complements to low-resolution observations, such as those with JWST, where multiple velocity components cannot be distinguished. }


\date{Received 7 September 2023 / Accepted
22 December 2023}

   \maketitle

%
%-------------------------------------------------------------------

\section{Introduction}\label{sec: intro}
The inner 10 au of protoplanetary disks are regions of high temperature and density, where snowlines of abundant molecules (H$_2$O and CO$_2$) and dust sublimation contribute to the conditions and chemistry. These regions may be the birthplaces of planets, whose properties and composition will be impacted by the conditions in the disk. 

High spectral resolution observations of infrared gas tracers are crucial probes for the kinematics and structure of the inner disk. Previous observations in the L ($\sim$3.5 \mic) and M band ($\sim$4.7 \mic) have shown that these wavelengths offer a unique view into the inner 10 au of disks, tracing the gas both inside and outside the dust sublimation radius, including from disk winds (see \citealt{banzatti22a,banzatti23a} and references therein). The molecular tracers include CO and H$_2$O (e.g., \citealt{najita03,blake04,brown13,banzatti15,banzatti17,banzatti22a}), and OH \citep{fedele11,brittain16}, whereas atomic tracers include Hydrogen recombination lines which can be used to determine the accretion rate \citep{salyk13, rigliaco15, komarova20}. The high spectral resolution that can be obtained for the emission from these tracers allow for kinematic analysis, including resolving multiple velocity components. This kinematic information can be useful in interpreting lower spectral resolution data, like that from JWST-NIRSpec and MIRI \citep{banzatti23a}. 

Most high spectral resolution L- and M-band studies of disks to-date have been done with VLT-CRIRES, Keck-NIRSPEC, and IRTF-iSHELL (see Table 1 of \citealt{banzatti22a} for an overview). In particular, VLT-CRIRES and Keck-NIRSPEC observations in the 2000s and early 2010s provided great new insight into the gas conditions and structure in the inner 10 au of protoplanetary disks, most powerfully by using the CO fundamental emission lines as a gas diagnostic. These results showed a wide array of line shapes, from the double-peaked line profiles that are typically associated with gas in Keplerian motion, to triangular line shapes associated to disk plus wind emission, and multiple absorption components with a range of blueshifts associated to absorption from a wind (e.g., \citealt{bast11,herczeg11,brown13,banzatti22a}). Upgraded Keck and VLT instruments became available recently \citep{martin18,dorn23}. With these high spectral resolution spectrographs available, now is a unique time to characterize the inner regions of protoplanetary disks. This is particularly true as observations by the James Webb Space Telescope (JWST) have larger spectral coverage but lower spectral resolution and the gas in the inner $\sim$10 au of disks are inaccessible to facilities like ALMA.

In this work, we present CRIRES+ Science Verification observations of the S CrA binary system using all of the available L- and M-band filter settings taken in 2021. The S CrA binary system is composed of two pre-main-sequence stars: S CrA N, a K7 star with stellar mass of 0.7 \Msun, and S CrA S, an M1 star with a stellar mass of 0.45 \Msun\ \citep{sullivan19}. The system is still somewhat embedded and has recently been found to have a large-scale streamer \citep{gupta23}. The disks in this system are borderline between Class I and II objects. The binary nature of this system complicates the distance determination from Gaia observations, therefore we adopt a distance of 150 pc, as used by \cite{sullivan19}. The binary separation in this system is 1.\arcsec3, such that both stars can be placed on the CRIRES slit at the same time. This system was also observed using CRIRES prior to the upgrade (oCRIRES), in 2007 Apr 22, 2007 Sep 3, and 2008 Aug 9, as is presented in \cite{brown13}. The binary nature of this system, paired with previous oCRIRES data, make S CrA an excellent test case for the upgraded CRIRES+.

We aim to provide new insight into the gas composition and conditions in the inner regions of the S CrA disks and characterize the upgraded CRIRES+ instrument with the goal of informing future observations, particularly with respect to spectral settings, to maximize efficiency and scientific output. The observations and data reduction are described in Section~\ref{sec: data reduction}. The results are presented in Section~\ref{sec: results} and are discussed, with an eye towards future observations and using CRIRES+ to complement other observations, in Section~\ref{sec: discussion}.

\begin{figure}[]
\includegraphics[scale=0.7]{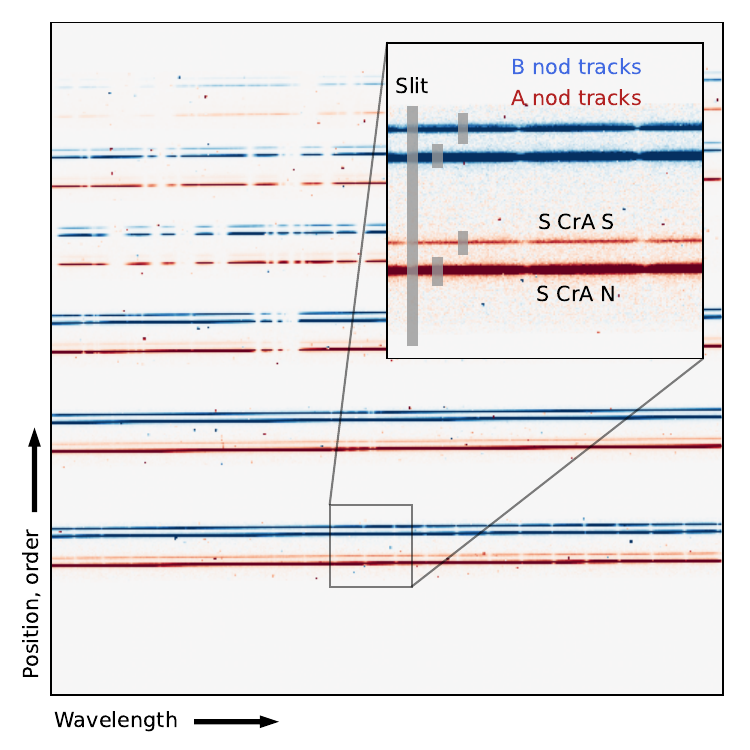}
\caption{Example of the detector image, showing the A (red) and B (blue) nod tracks for S CrA N (bottom track) and S CrA S (top track). The slit is illustrated on the left in the inset, although in reality the slit is tilted. The small grey boxes correspond to the regions where the spectra were extracted for each source and each nod position. 
     \label{fig: detector}}
\end{figure}

\begin{figure*}
\includegraphics[scale=0.66]{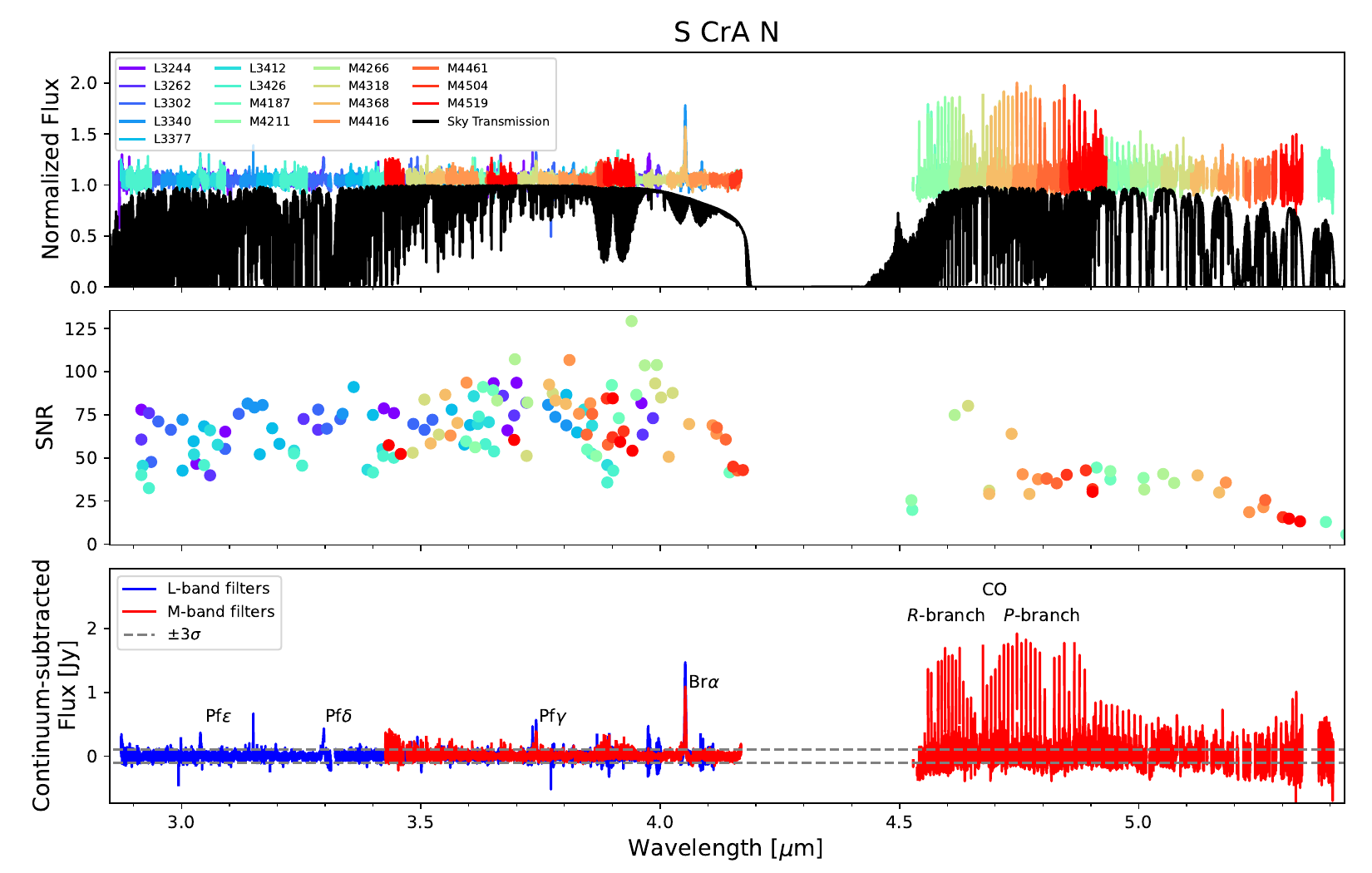}
\caption{Top: S CrA N CRIRES+ reduced and telluric-corrected spectra in all available L- and M-band settings (colors). The wavelength separation of the orders can be seen; for instance, the M4519 (red) filter shows five orders (two orders fall in the telluric absorption gap from 4.2 to 4.5 \mic). Note that between 3.4 and 4.2 \mic\ L- and M-band filters overlap. The sky transmission is shown in black, for reference. Middle: The signal-to-noise ratio for each filter, order, and detector in the S CrA N spectra. Bottom: The combined L- (blue) and M-band (red) spectra for S CrA N after flux calibration and continuum subtraction. Prominent emission features are labeled. Regions where the atmospheric transmission is below 50\% are masked. The gray lines indicate the $\pm$3$\sigma$ regions for reference. }
     \label{fig: reduced}
\end{figure*}

\section{Observations and data reduction} \label{sec: data reduction}
\subsection{Observations}\label{subsec: observations}
The VLT-CRIRES+ observations that we present here were taken during the Science Verification phase as part of Program 107.22T7 (PI A. Bosman). The data were taken on 18 and 19 September 2021. All L- and M-band spectral settings (seven settings in L band and nine in M band) were utilized, which given the upgraded cross-dispersed nature of CRIRES+, provides full coverage from $\sim$2.9--5.5 $\mu$m. The slit was placed such that both components of the S CrA binary system were oriented on the slit, which was set to the 0.$''$2 ($R\sim$80000) slit width (Figure~\ref{fig: detector}). The adaptive optics (AO) system was employed during the observations. Two position angles were observed in the L3340 and M4368 filters, with an offset of 180$^{\circ}$; however, a manual shift was introduced after the change in position angle and this caused problems with guiding, resulting in the data not being suitable for spectroastrometry. Each filter was observed with an integration time of 10 seconds for three exposures per nodding position and with two ABBA nodding patterns for a total integration time of 240 seconds per filter per position angle. The nod throw was 5 arcseconds along the slit. In addition to S CrA, a telluric standard star, $\lambda$ Aquilae ($\lambda$ Aql, a B9V star with a V-band magnitude of 3.43), was observed for telluric correction with an integration of 10 seconds per exposure with two exposures per nod and two ABBA nodding cycles for a total integration time of 160 seconds. We note that metrology, which is used to fine-tune the placement of the gratings to maximize wavelength calibration, was not available in the L and M band during these observations; however, this is now available at these wavelengths and should be used, per ESO recommendations. An observing log is provided in Table~\ref{tab: obs log}.

\subsection{Data reduction and calibration}\label{subsec: datareduction}

We utilized the ESO CR2RES EsoReflex pipeline for initial data reduction of the 2D images (\citealt{reflex, dorn23}; see Figure~\ref{fig: detector}) using version 1.0.5 \footnote{https://www.eso.org/sci/software/pipelines/crires/crires-pipe-recipes.html}. The default spectral extraction in the CR2RES pipeline can handle arbitrary slit functions by simply taking all of the flux along the slit. For S CrA, this would result in a single, combined spectrum for both components of the binary. However, to extract each component of the binary separately, we use the EsoRex reduction method on the reduced combined A and B frame images. To do this, we use the \texttt{slit$\_$frac} keyword in the \texttt{cr2res$\_$util$\_$extract} function, to specify the regions where the spectra for each star should be extracted. This is presented in Figure~\ref{fig: detector}. As the seeing conditions were fairly good throughout the observations (Table~\ref{tab: obs log}), AO was used, and the targets have fairly bright continua, the traces for each source are well-separated, resulting in negligible contamination. This produced an A nodding position and a B nodding position spectrum for each star, in each chip, for all of the observed filters.

The CRIRES+ slit is tilted on the detector projection, therefore to combine the extracted A and B nod position spectra, we first determine the offset between a given A and B nod pair using the Astropy Specutils \texttt{template$\_$correlation} program. If a good correlation was not found, we identified the shift between the A and B frame spectra by eye. After finding the shift between the spectra, we shift the B spectra to align with the A spectra before combining. This was done for each detector, in each order, for all filters, for S CrA N, S CrA S, and the telluric standard star separately.

There is no lamp available for use in the L and M band to use for wavelength calibration. Therefore, telluric sky lines were used to wavelength calibrate the spectra, as was done for oCRIRES data. To do this, we use the telluric-correction tool \texttt{Molecfit} \citep{kausch15,smette15}. \texttt{Molecfit} performs synthetic modeling of telluric lines. The best-fit \texttt{Molecfit} model provides a wavelength solution that is free from any overall wavelength shifts, stretches and/or compressions of the spectra due to the slit tilt. Therefore, even if the spectra are not telluric-corrected using \texttt{Molecfit} (see below), using \texttt{Molecfit} is still necessary to get a proper wavelength calibration in the L and M band. 

The correction of telluric sky lines can be done with \texttt{Molecfit} or using a telluric standard star if the standard star was observed close in time and airmass (typically with a difference in airmass up to 0.2; \citealt{ulmermoll19}), and in similar observing conditions. For the S CrA observations, $\lambda$ Aql was observed as the telluric standard. The L-band observations were taken on the night of 18 September 2021 and the difference in airmass between S CrA and $\lambda$ Aql was greater than 0.2 (S CrA had an airmass of 1.245 to 1.494 while $\lambda$ Aql had an airmass of 1.857 to 2.881). The M-band data, taken the following night, were observed within an airmass difference of $\sim$0.2 (S CrA had an airmass of 1.023 to 1.047 and $\lambda$ Aql had an airmass of 1.104 to 1.268). Therefore, the L-band data are telluric corrected using \texttt{Molecfit} and the M-band data are corrected using the telluric standard star. Although the telluric star was not used to telluric correct the L-band data, we still use this data to correct for instrumental effects (blaze function, fringing, etc.) in the data. Alternatively, the blaze function can be corrected by dividing the reduced spectrum by the blaze spectrum, which is a product of the calibration cascade. Finally, early A-type and late B-type stars that are largely featureless and therefore useful for correcting for telluric lines, can have photospheric absorption in the hydrogen lines. We fit this absorption using a quadratic spline interpolation in the $\lambda$ Aql spectra where photospheric absorption lines are detected and remove it before using those spectra to telluric correct the science target spectra. We find that for our sources higher quality data are achieved when using a telluric standard star to correct for telluric lines, however, for targets with high signal-to-noise on the continuum, \texttt{Molecfit} can provide a high quality telluric correction.

WISE Band 2 photometry are typically used for flux calibration of M-band data, however, the resolution of WISE is not sufficient to distinguish each component of the binary. \cite{sullivan19} provide photometry for S CrA N and S. Using Keck II/NIRC2, the authors determine that the 3.45 \mic\ flux 2.11$\pm$0.07 Jy for S CrA N and 0.72$\pm$0.4 Jy for S CrA S.

All of the spectral settings were combined into a single spectrum for each target, taking the weighted average in regions where multiple filters overlapped. This is done taking the A and B nod spectra and computing the weighted average at each wavelength point, $i$, as \begin{equation}
   F_i = \frac{(F_{i,A}/\sigma_{i,A}^2)+(F_{i,B}/\sigma_{i,B}^2)}{(1/\sigma_{i,A}^2)+(1/\sigma_{i,B}^2)}
\end{equation}
with an associated error given by \begin{equation}
    \sigma_i = \sqrt{\frac{1}{\sigma_{i,A}^2}+\frac{1}{\sigma_{i,B}^2}}. 
\end{equation}
The errors at each point for the A and B frames are the errors from the pipeline.

The S CrA N spectra in all filters and after combining all filters is presented in Figure~\ref{fig: reduced}. Additionally, the final L- and M-band spectra are presented in the Appendix in Figures~\ref{fig: overview Lband} and \ref{fig: overview Mband}, respectively, for S CrA N, and in Figures~\ref{fig: overview Lband S CrA S} and \ref{fig: overview Mband S CrA S} for S CrA S. The data from oCRIRES are shown for comparison and the coverage of each CRIRES+ L- and M-band spectral setting is shown for reference. The combined spectra are available for both targets on \url{spexodisks.com}.

\section{Results}\label{sec: results}

\subsection{Signal-to-noise ratio}
We estimated the signal-to-noise ($S/N$) ratio from the final spectra by identifying a region of continuum in each detector, order, and filter. We then estimated the $S/N$ in each of the spectra by determining the dispersion in the selected region. All filters were observed with the same integration time (240 total seconds), therefore the dispersion in $S/N$ ratio observed depends on the sensitivity of the filter, the flux of the sources which changes as a function of wavelength from 3 to 5.5 \mic, and proximity of the continuum region to strong and/or numerous telluric features, which results in residuals that lower the $S/N$ and where there is simply less signal due to the atmospheric absorption. For S CrA N, the highest $S/N$ of $\sim$100 is achieved from 3.5 to 4.0 \mic. The highest $S/N$ red-ward of the strong break at 4.3 \mic\ due to telluric absorption, is $\sim$70 around 4.6 \mic. The distribution of $S/N$ values in the S CrA N spectra is presented in Figure~\ref{fig: reduced}. The signal-to-noise in the final spectrum depends on how many overlapping filters are combined, however, we reach a noise level of $\sim$0.03 Jy at 4.9 \mic\ in the S CrA N spectra.

In comparison to the oCRIRES data of S CrA N, after taking into account the difference in integration time between observations, we find a sensitivity increase of $\sim$10\%. The integration time for a given filter in the new CRIRES+ data are shorter than the oCRIRES data, resulting in overall lower $S/N$, despite the increase in instrument sensitivity between observations.

\subsection{Line detections}
Across the L- and M-band spectra numerous ro-vibrational emission lines of $\mathrm{H_2O}$, OH and CO and recombination lines of atomic hydrogen were detected (Figure~\ref{fig: overview Lband}, \ref{fig: overview Mband}, \ref{fig: overview Lband S CrA S}, and \ref{fig: overview Mband S CrA S}), particularly in S CrA N, where the brighter nature of this source resulted in higher $S/N$ spectra. 

\subsubsection{Hydrogen}
\begin{figure*}[]
\centering
\includegraphics[scale=0.5]{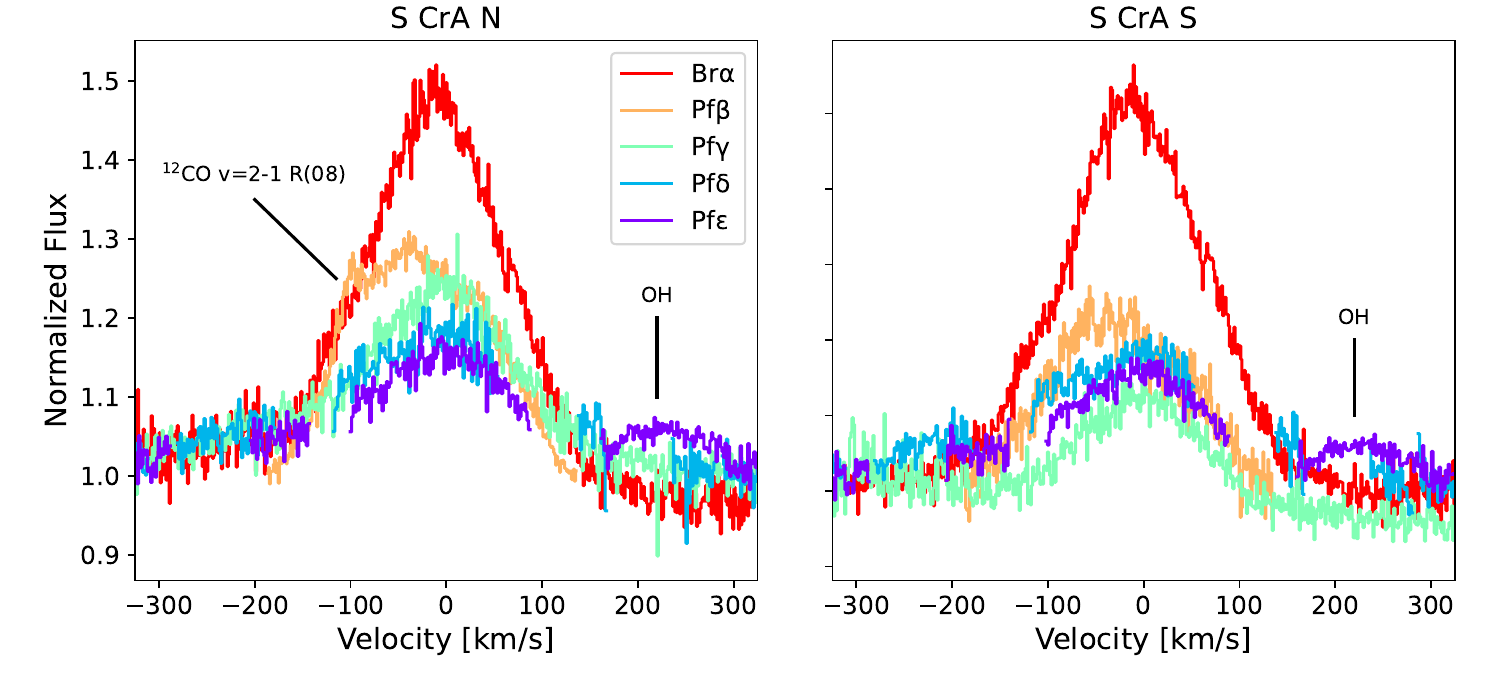}
\caption{Hydrogen recombination lines detected in the S CrA N (left) and S CrA S (right) spectra. A $^{12}$CO v=2-1 line is blended with the Pf$\beta$ feature in S CrA N and an OH doublet is blended with Pf$\epsilon$ for both targets.} 
     \label{fig: hydrogen}
\end{figure*}

We detect Hydrogen recombination lines in the Brackett and Pfund series in both S CrA N and S (Figure~\ref{fig: hydrogen}). Br$\alpha$ has broad wings. Pf$\gamma$ and Pf$\delta$ show an additional blue-shifted emission in S CrA N. Pf$\beta$ is blended with the $^{12}$CO v=2-1 R(8) line in S CrA N. Pf$\epsilon$ is blended with an OH doublet for both targets.

 \subsubsection{OH}
 Several OH lines are detected in the L band around 2.93 \mic\ in the S CrA N spectrum, also seen in the oCRIRES data analyzed in \cite{banzatti17}. The lower $S/N$ ratio in our data relative to those in \cite{banzatti17}, make a comparison difficult, particularly for these weak features. However, we see no qualitative differences in the features.

\subsubsection{H$_2$O}
In the L band of S CrA N, we detect the same water lines previously observed by \cite{banzatti17}. Given the low $S/N$ of the spectrum we did not analyse the L-band water lines, limiting to highlight the detections. In the M band we detected $\mathrm{H_2O}$ emission lines from the ro-vibrational $R$-branch of the bending mode ($\nu_2$), as observed in other disks \citep{banzatti22a,banzatti23a}. These M-band lines are faint, at the $\sim$5\% continuum level. The lines are fairly broad as shown in the last panel of Figure~\ref{fig: stacked}. No H$_2$O emission is detected in the S CrA S spectrum.

\subsubsection{CO}\label{subsec: CO detection}
\begin{figure*}[]
\includegraphics[width=0.99\textwidth]{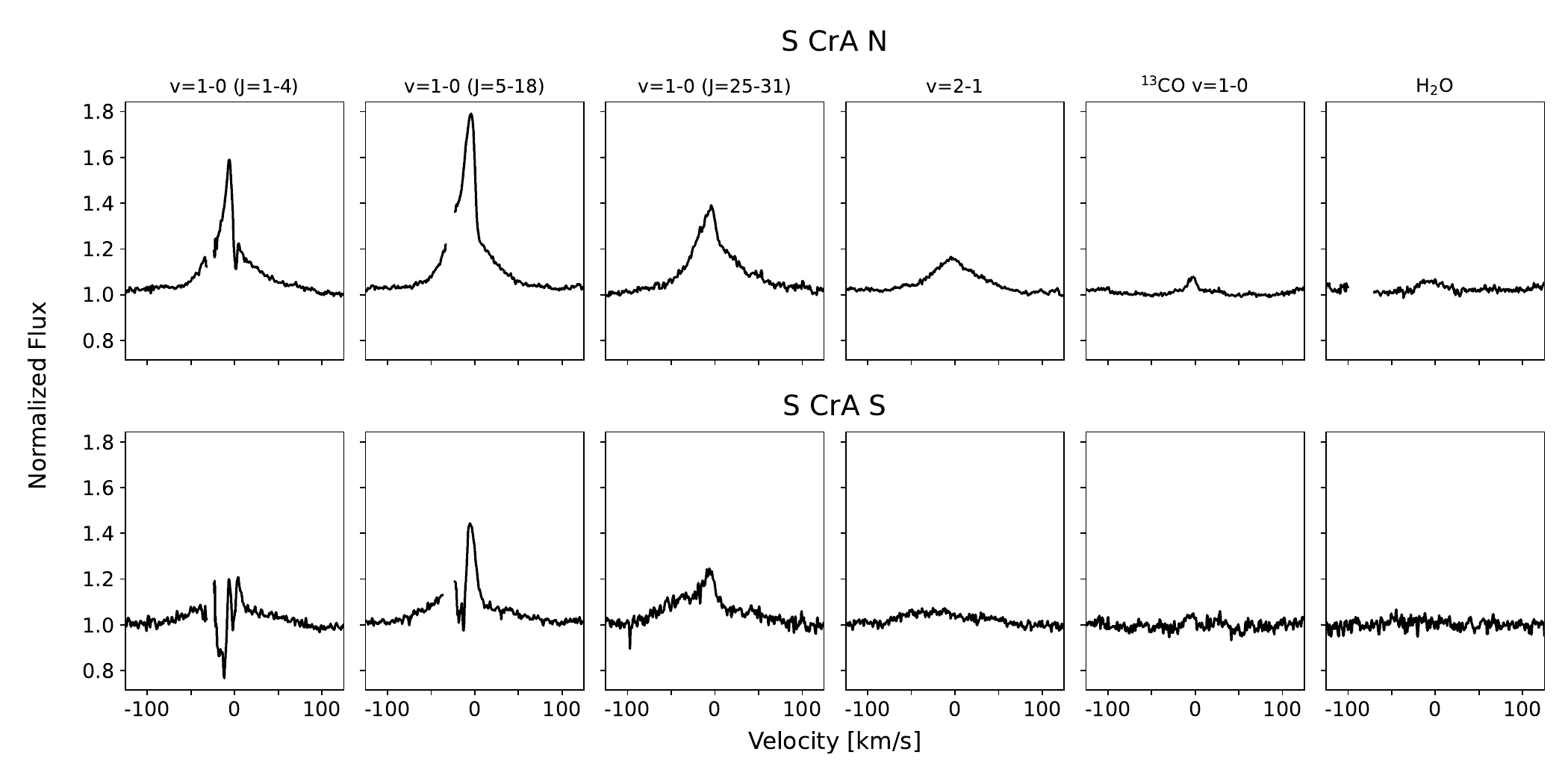}
\caption{Stacked $^{12}$CO, $^{13}$CO, and H$_2$O line profiles for S CrA N (top) and S CrA S (bottom). Blank portions correspond to masking due to telluric lines in the case of the first two panels and the masking of another feature in the case of the H$_2$O line profile in S CrA N.
     \label{fig: stacked}}
\end{figure*}

In S CrA N, we detect $^{12}\mathrm{CO}$ emission lines in the v=1-0, v=2-1 and v=3-2 ro-vibrational branches. $^{13}\mathrm{CO}$ is instead detected only in the v=1-0 branch. In S CrA S, only the $^{12}$CO v=1-0 lines are detected. We analyzed these lines following the procedure described in \cite{banzatti22a}. Two velocity components are detected in the $^{12}$CO v=1-0 and v=2-1 lines. In the v=3-2 lines only a broad component is observed, while in the $^{13}$CO v=1-0 lines only the narrow component is present. The broad component (BC) has a full width at half maximum (FWHM) of 72 km s$^{-1}$ and the narrow component (NC) has a FWHM of 14 km s$^{-1}$. In S CrA S, the BC and NC v=1-0 FWHM are 126 km s$^{-1}$ and 10 km s$^{-1}$, respectively, and multiple absorption components are detected at different blue-shifts, especially in the low $J$ lines. These velocity components are discussed in terms of emitting radii in Section~\ref{subsec: emitting radii}. The stacked CO and M-band H$_2$O line profiles are presented in Figure~\ref{fig: stacked}. 

\subsection{Other species}\label{subsec: other species}
The full spectral range in these observations, from $\sim$2.9 to 5.5 \mic, covers many additional molecular transitions, including those from H$_2$, C$_2$H$_2$, HCN, NH$_3$, CH$_4$, and SO$_2$ \citep{mandell12}. Emission from these features is not detected in our data above a level of 3$\sigma$.

\subsection{Rotational diagrams}

With sufficient coverage of multiple ro-vibrational lines, it is possible to use a rotational diagram as a diagnostic tool to study the temperature, density, and excitation processes of the gas producing the observed spectrum. The rotational diagrams, also known as Boltzmann diagrams, show the line flux versus the excitation energy or upper-level quantum number. The quantity on y-axis is
\begin{equation}
    y = \ln \left( \dfrac{4 \pi F_{ul}}{h\nu_{ul} g_{u} A_{ul}} \right) \label{eq:y_rot}
\end{equation}
where $F_{ul}$ is the flux of the line at frequency $\nu_{ul}$, produced by the transition from the upper level $u$, which has statistical weight $g_{u}$, to the lower level $l$. Finally, $A_{ul}$ is the Einstein coefficient associated to the transition.
If the gas is in local thermodynamic equilibrium (LTE) and the lines are optically thin, the rotational diagram results in a straight line with a slope given by $-1/T_\mathrm{kin}$ and the intercept is proportional to the total mass of the gas. 
Different combinations of gas temperature and density can result in a deviation from LTE, whereas optical depth effects can produce distinct curvatures in the rotational diagram (e.g., \citealp{goldsmith99}). Moreover, different excitation processes such as UV- and IR-pumping can introduce a deviation from a simple straight line in the rotational diagram up to $E_\mathrm{up}\sim7000~\mathrm{K}$ \citep{thi13}. Thus, line coverage up to high $E_\mathrm{up}$ values is necessary to identify such deviations and in general, it is necessary to cover both low-$J$ and high-$J$ lines to accurately constrain the gas conditions.

\begin{figure*}[]
\includegraphics[width=0.98\linewidth]{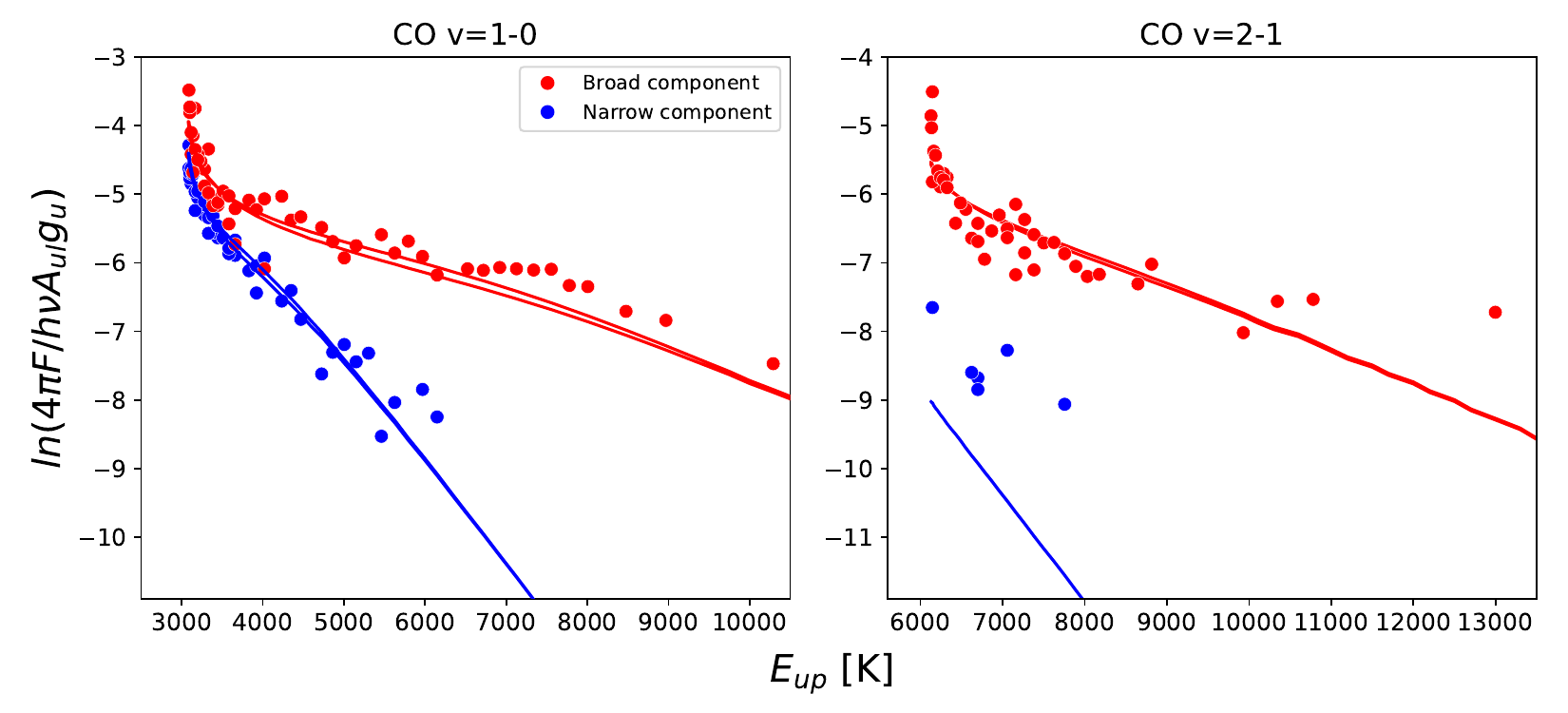}
\caption{Rotational diagrams of $^{12}$CO v=1-0 (left) and $^{12}$CO v=2-1 (right) for S CrA N. The broad and narrow components are shown in red and blue, respectively. The data are shown as the points and the best-fit model is shown in the lines. The model is fit for $^{12}$CO v=1-0 and predicted for the v=2-1 transitions. The two model lines correspond to the $P$- and $R$-branch lines. 
     \label{fig:rotdiag all}}
\end{figure*}

The rotation diagram comparing the v=1-0 and v=2-1 broad and narrow components in S CrA N is presented in Figure~\ref{fig:rotdiag all}. By being able to distinguish these two velocity components, the rotation diagram can then be used to determine the conditions probed by these two different velocity components (which correspond to different radii in the disk, see Section~\ref{subsec: emitting radii}). For each component, we fit the rotational diagram of the v=1-0 lines with a single-slab model in LTE, following the fit procedure of \cite{banzatti12}. This is done after flux calibration. As there is no separated photometry at 4.7 \mic\ for this system, we use the 3.45 \mic\ photometry from \cite{sullivan19} given that the S CrA N spectral energy distributions is relatively flat between 3.45 and 8 \mic, the closest photometry points. We discuss the impact of flux uncertainty on the model parameters below. The model depends on only three parameters, namely the kinetic temperature of the gas $T_\mathrm{kin}$, the column density $N_\mathrm{col}$, and the emitting area $A$. For each model, the line fluxes are calculated and compared to the measured line fluxes by calculating the $\chi^2$. The best-fit parameters are then found by a $\chi^2$ minimization. 

The temperature and column density of the best-fit model for the BC lines are $T_\mathrm{kin, BC} = 1820\pm180~\mathrm{K}$ and $N_\mathrm{col, BC} = (3.5\pm1.5)\times10^{17}~\mathrm{cm}^{-2}$, while for the NC lines they are $T_\mathrm{kin, NC} = 640\pm80~\mathrm{K}$ and $N_\mathrm{col, NC} = (8.0\pm3)\times10^{16}~\mathrm{cm}^{-2}$. Uncertainties are the 5$\sigma$ uncertainties from the model grids. The emitting area is 0.24 au$^{2}$ for the broad component and 4.19 au$^{2}$ for the narrow component. The absolute flux level only impacts the emitting area. The flux uncertainty at these wavelengths is $\sim$5\%, resulting in uncertainties on the emitting area of 0.01 and 0.21 au$^{2}$ for the broad and narrow components, respectively.

\begin{figure}[]
\includegraphics[width=0.98\linewidth]{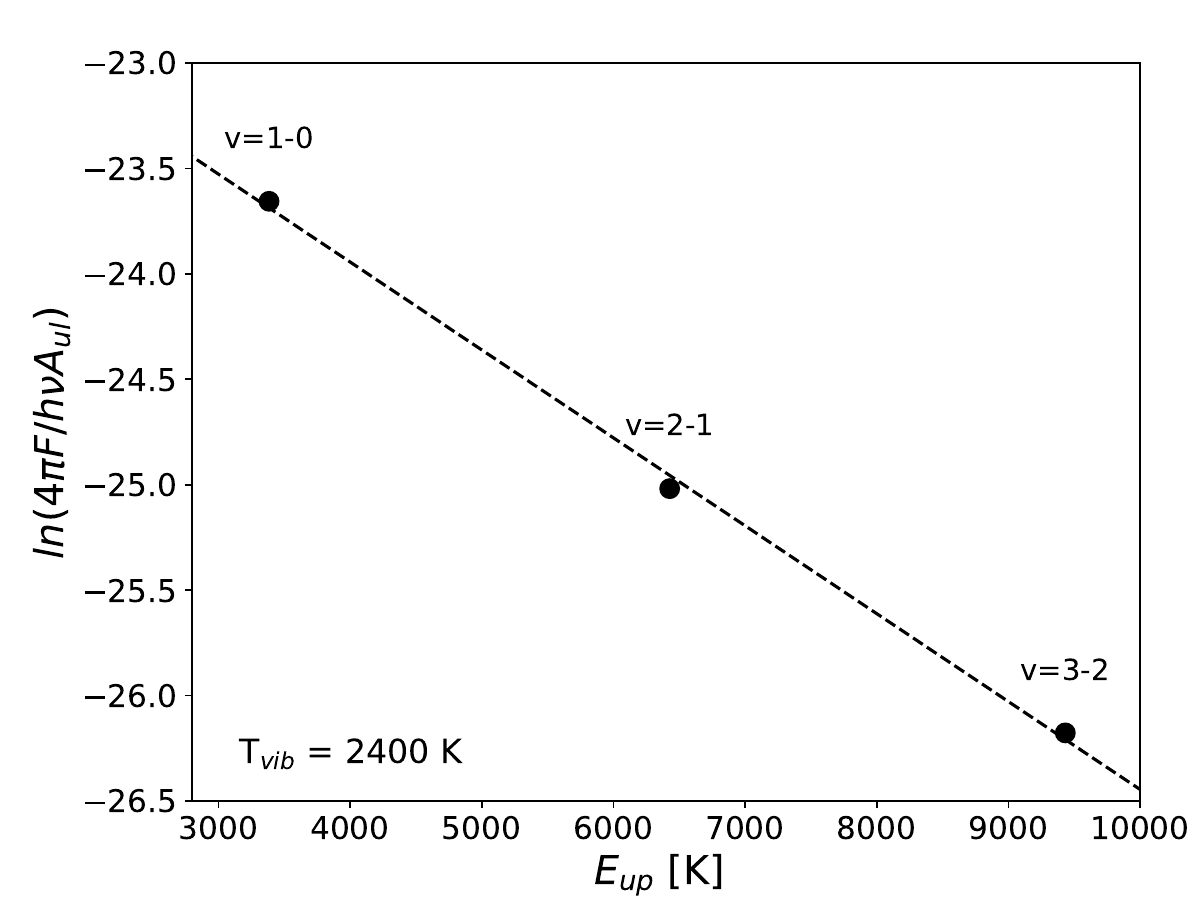}
\caption{Vibrational diagram of the $^{12}$CO broad component using the R(09) line in S CrA N. The slope indicates a vibrational temperature of 2400 K.
     \label{fig:vibdiag}}
\end{figure}

The v=2-1 BC lines are generally well reproduced, with the exception of the high energy lines $E_\mathrm{up}\gtrsim 10^4~\mathrm{K}$. These results suggest that the BC emission is close to LTE, both rotationally and vibrationally. This is also seen in the vibrational diagram for $^{12}$CO in S CrA N, done using the broad component of the R(9) line, which is presented in Figure~\ref{fig:vibdiag}. This is not done for the narrow component, as there is no narrow component to the R(9) line in the v=2-1 and v=3-2 transitions. A linear fit to the points results in a vibrational temperature of 2400 K, higher than the rotational temperature of 1820 K, as has been found in Herbig Ae/Be stars (e.g., \citealt{brittain07a, vanderplas11,banzatti22a}) and even in some T Tauri stars (e.g., \citealt{bast11}). This may be evidence of UV pumping, which could then be expected to be present in S CrA N. In the case of the NC emission, the best-fit model is not able to reproduce well the v=2-1 rotational diagram. This suggests that the NC emission is rotationally but not vibrationally thermalized. The gas producing the NC emission could have lower density, lower than the critical density of the CO lines and lower than the gas traced by the BC.

In previous disk samples observed with high spectral resolution in the M band, most targets did not have coverage up to the high $E_\mathrm{up}$ values that we cover here (e.g. \citealp{najita03}, \citealp{blake04}, \citealp{brown13}). Recently, the iSHELL spectrograph \citep{rayner16,rayner22} at the NASA Infrared Telescope Facility (IRTF) has made it possible to get a complete coverage of v=1-0 lines up to $E_\mathrm{up}\sim$10,000$~\mathrm{K}$, thanks to the covering of the entire M band (from 4.52 to 5.24 \mic) in one single shot. This has allowed to get the most complete coverage of the rotational diagram to-date, revealing the full shape of curvatures associated to different gas conditions \citep{banzatti22a}.

\begin{figure*}[]
\includegraphics[scale=0.58]{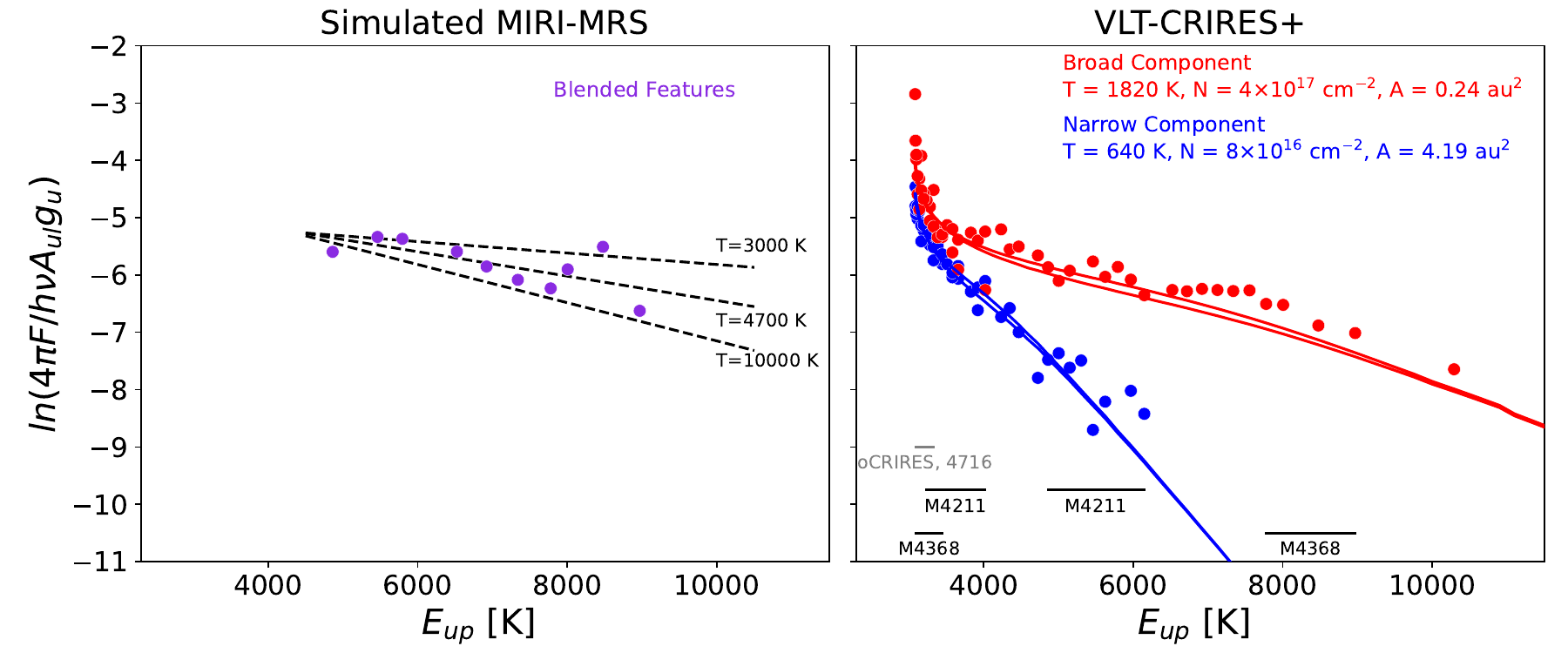}
\caption{Rotational diagrams of the $^{12}$CO v=1-0 transitions. On the left is a simulation of the S CrA N CRIRES+ data with the spectral resolution and wavelength coverage of JWST MIRI-MRS, which does not have the resolution to distinguish between different kinematic components and only covers high-$J$ lines. Three optically thin models with different temperatures are shown for illustration. The 4700 K is the best linear fit result. On the right is what is observed with VLT-CRIRES+. Only detected transitions are shown. The lines correspond to a linear fit to the simulated MIRI data (left) and the best-fit LTE model to the CRIRES+ data (right). In order to constrain the gas temperature and column density, sufficient upper energy level coverage is needed. In the right panel, we provide the coverage of two M-band filters of CRIRES+ (black) that provide sufficient to characterize the CO gas. The coverage of one of the oCRIRES filters is shown in grey. The upgrades to CRIRES+ result in greater energy level coverage with a single filter. The two model lines correspond to the $P$- and $R$-branch lines. } \label{fig: crires miri}
\end{figure*}

As CRIRES+ does not cover the entire M band in a single spectral setting, it is desirable to determine what filters are needed to cover sufficient CO lines to achieve good coverage for CO analysis while maximizing observing efficiency. This is presented in Figure~\ref{fig: crires miri}, where the v=1-0  rotational diagram is plotted in the right panel. The upper energy coverage by filters M4211 and M4368 are indicated, in comparison with one filter from oCRIRES, which had a shorter simultaneous wavelength coverage. This demonstrates that even with just two settings it is possible to sufficiently cover the v=1-0 rotational diagram from low energy levels ($E_\mathrm{up} \sim 3000~\mathrm{K}$) to high energy levels ($E_\mathrm{up} \sim 8000~\mathrm{K}$). M4318 covers a similar range of CO lines as the M4368 filter, however, M4368 also covers the Br$\alpha$ line at 4.05 \mic, which offers an isolated line for determining the simultaneous accretion rate, and H$_2$O lines at 5 \mic.

\section{Discussion}\label{sec: discussion}

\subsection{CO emitting radii}\label{subsec: emitting radii}
With sufficient spectral resolution, Keplerian broadening of lines can be observed and used to determine the emitting radius of the gas. Under the assumption of Keplerian rotation, the emitting radius $R$ depends on the line width $\Delta$v through
\begin{equation}
    R = M_* G {\left(\dfrac{\sin i}{\Delta \mathrm{v}}\right)} ^2 
\end{equation}
where $M_*$ is the mass of the central star and $i$ is the inclination of the disk. Additionally, as in S CrA N, if multiple velocity components are observed in the lines, an emitting radius can be determined for each component. We determine the CO emitting radius for the broad and narrow velocity components in S CrA N and S using inner disk inclinations from VLTI-Gravity \citep{gravity21}.

\begin{figure*}[]
\includegraphics[scale=0.7]{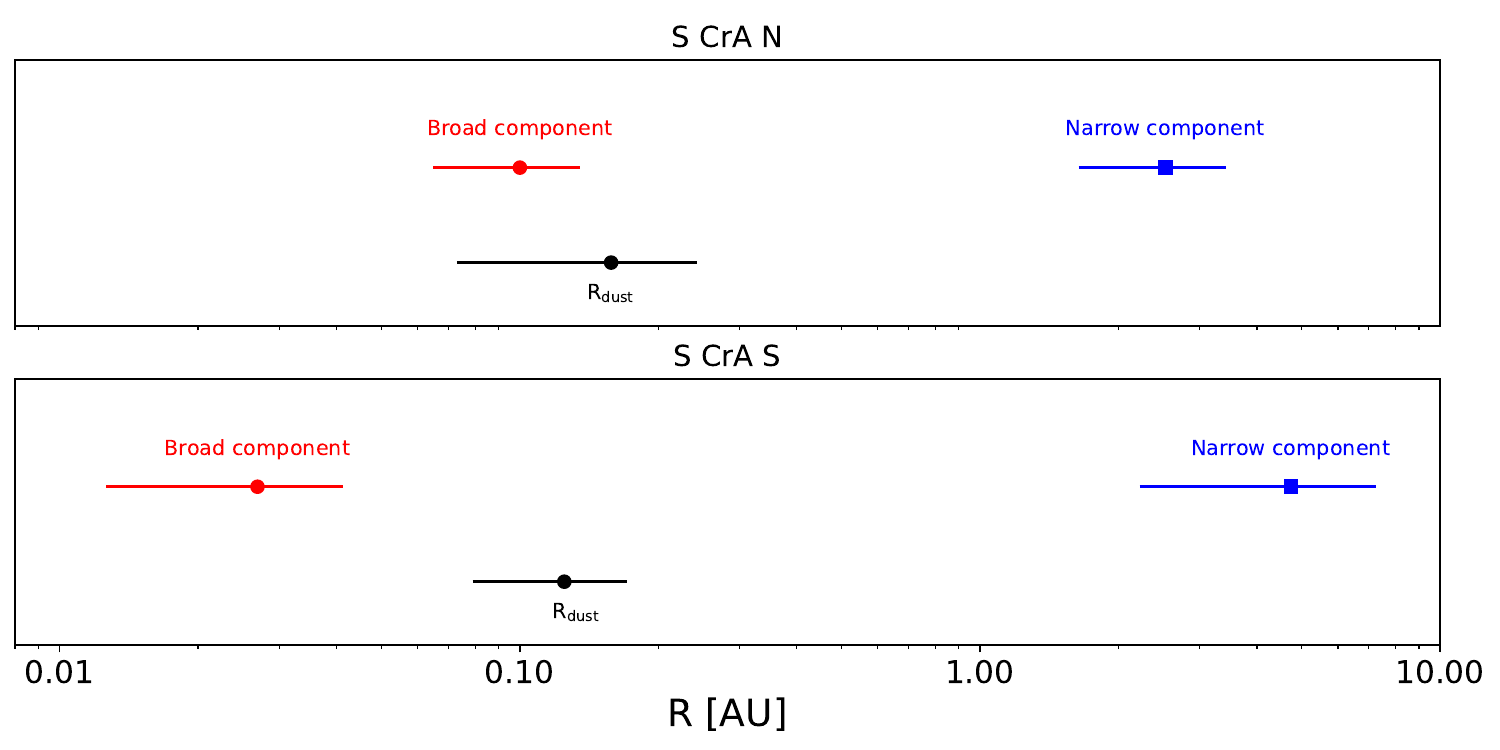}
\caption{Emitting radii of the CO lines (colors) compared to the dust continuum radius determined from VLTI-Gravity (black circle; \citealt{gravity21}, values corrected for a different adopted distance). S CrA N is presented in the top panel and S CrA S in the bottom panel. For both stars, the broad component is coming from the inner, dust-free region, while the narrow component is coming from larger disk radii.
     \label{fig: emitting radii}}
\end{figure*}

The CO radii for the broad and narrow components are presented in Figure~\ref{fig: emitting radii}. The emitting radii can be put into context using information on the disk structure from other methods, for instance from near-infrared interferometry or from the dust sublimation radius, estimated given the stellar luminosity. For S CrA N and S, the broad component is coming from inside the dust disk rim, as determined from VLTI-Gravity observations \citep{gravity21} and as found in other highly accreting disks (Figure 12 in \citealt{banzatti22a}). The narrow component is coming from farther out, beyond $\sim$1 au in both disks. 

The two velocity components seen in S CrA N and S are not uncommon in protoplanetary disks \citep{banzatti22a}. Having high spectral resolution means that not only can the gas properties be studied for these components separately, but it also provides a map of the inner disk structure. Paired with information on the dust distribution (e.g., from interferometry or a determination of the dust sublimation radius), a more complete picture of the inner 10 au of protoplanetary disks can be accessed. This is useful information on its own, however, it can also be a crucial complement to low spectral resolution data, like that from JWST-MIRI, where determining the emitting area, is dependent on the model and on whether the gas is optically thin or optically thick and the actually emitting radius, is unconstrained. In this way, high spectral resolution spectroscopy is a uniquely powerful tool.

\subsection{Lessons learned for JWST observations}
The high spectral resolution and large spectral coverage of these CRIRES+ observations allow for a detailed characterization of the CO gas in the S CrA disks. The spectral resolution allows for the extraction of multiple velocity components, which come from different regions in the disk, and the large coverage of the CO lines allows for access to the full CO ro-vibrational ladder for each velocity component. It is useful then to consider how this information can be used and what it would mean if these observational features were not available. This is particularly useful to consider as JWST is providing great insight into disk chemistry, but is lacking the spectral resolution to distinguish many line blends, let alone distinguish different velocity components in a single feature. Additionally, JWST-MIRI MRS observations lack full coverage of the whole CO ladder. In Figure~\ref{fig: crires miri} we show what the $^{12}$CO rotational diagram looks like with CRIRES+ and simulate what this rotation diagram would look like given the spectral resolution and coverage of JWST-MIRI MRS. To do this, we bin the CRIRES+ resolution to that of MIRI at 5 \mic\ (thereby losing the ability to distinguish between velocity components), along take the spectrum long-ward of the MIRI MRS wavelength limit of 4.9 \mic, and re-extract the fluxes for the covered CO lines. The curvature of the rotation diagram around upper energy levels of $\sim$4000 K is lost, resulting in non-convergence of the models. A simple straight line fit to the rotation diagram, assuming optically thin emission, results in a very high temperature model (between $\sim$ 3000 K and 10000 K), despite the fact that the broad component, that dominates the flux, is actually better fit by a lower temperature and higher column density when the full energy level coverage is considered in the fit. 

JWST-NIRSpec observations would provide broader coverage of the CO ladder, including avoiding the gap in ground-based data caused by the strong telluric absorption short-ward of 4.5 \mic, which would allow for higher $R$-branch lines to be detected. However, the lower spectral resolution of JWST-NIRSpec relative to MIRI, will make for even more severe line blending. Observations from VLT-CRIRES+ and other high spectral resolution ground-based spectrographs are crucial complements to observations from facilities like JWST.

\subsection{Comparison with other sources}\label{subsec: comparison to other sources}
Large samples of disks have been studied using CO ro-vibrational emission, including in \cite{herczeg11}, \cite{brown13}, and \cite{banzatti22a}. These surveys have covered disks at different evolutionary stages and different stellar mass regimes. Much has been learned about the inner disk gas in disks as a result of these stellar and evolutionary properties. \cite{brown13} find that the CO line profile complexity decreases with increasing evolutionary state due to the disappearance of ice features and foreground absorption. However, those authors find that the underlying emission is quite similar between Class I and Class II disks. No ice absorption is observed in the L or M band in either S CrA N or S, but some CO absorption is present in the low-$J$ lines of S CrA N and in the low- to intermediate-$J$ lines in S CrA S, indicative of some envelope and/or cloud absorption \citep{herczeg11}. The lack of ice features, paired with evidence of some remnant envelope material, indicates an intermediate evolutionary state for this binary system, which agrees with analysis of the spectral energy distributions that determine these are flat-spectrum objects in the transition phase between Class I and II \citep{sullivan19}.

\section{Summary and conclusions}\label{sec: conclusions}
This work presents VLT-CRIRES+ high-resolution ($R$=80000) spectroscopy of the binary system S CrA. A full spectral scan in the L- and M-band filters allows for a detailed analysis of the upgraded instrument for studying protoplanetary disks and a unique look into this binary system. For the characterization of the upgraded CRIRES+ instrument, we find:

\begin{itemize}
    
    \item The sensitivity of CRIRES+ is $\sim$10\% better than oCRIRES, determined by comparing previous observations of this binary system to the data presented here. 

    \item We find that with only two CRIRES+ spectral settings, the CO rotational diagram can be sufficiently covered. M4211 and M4368 provide good CO energy level coverage, and M4368 additionally covers the accretion-tracing Br$\alpha$ line at 4.05 \mic\ and H$_2$O lines at 5 \mic. Many more lines were covered simultaneously with CRIRES+, compared to the smaller spectral coverage of oCRIRES, which allows for a more detailed analysis of the CO emission lines.

\end{itemize}

For the S CrA binary system, we find: 
\begin{itemize}
    \item Atomic (hydrogen recombination) and molecular (OH, H$_2$O, and CO) lines are observed in the spectrum of S CrA N. In the fainter S CrA S disk, only the $^{12}$CO v=1-0 and hydrogen recombination lines are detected.

    \item The high spectral resolution of CRIRES+ (in particular with the 0.\arcsec2 slit), allows for the detection of multiple velocity components in the CO lines observed in these disks. A narrow, low-velocity component and a broad, high-velocity component are detected in both disks. 

    \item Rotation diagrams are important tools for determining the gas temperature, column density, and determining optical depth effects and/or excitation processes. For S CrA N, we find that the broad and narrow velocity components of the CO emission are coming from gas with similar column densities ($\sim$8$\times$10$^{16}$-4$\times$10$^{17}$ cm$^{-2}$), but different temperatures (1820 K and 640 K, for the BC and NC, respectively) and emitting areas (0.24 au$^2$ and 4.19 au$^2$, respectively). 
    
    \item There is a clear separation in the CO emitting radii between the broad component ($\sim$0.1 au and $\sim$0.03 au in S CrA N and S, respectively) and the narrow component ($\sim$3 au and $\sim$5 au, respectively). The broad components are coming from within the dust-free inner region, where the inner dust disk radii were determined from VLTI-Gravity observations. The narrow component may be tracing a wind at larger radii (see \citealt{banzatti22a} for examples of disks winds observed in CO line profiles in other sources), in agreement with the emitting area distinction found in fitting the rotational diagrams. 
\end{itemize}

We are now in a time when several high spectral resolution ($R\gtrsim$75000) spectrographs capable of observing in the L and M band are online and the future for such observations is bright, in particular with E-ELT METIS achieving first light in the late 2020s. VLT-CRIRES+, in the years before METIS, is the only telescope-instrument combination that can observe targets in this wavelength range with this resolution at declinations below $\sim$-40$^{\circ}$. VLT-CRIRES+ has then great potential for continuing the work of oCRIRES and producing new, cutting-edge results.

\begin{acknowledgements}
We thank Alexis Lavail and the ESO User Support team for their help in reducing this very early CRIRES+ data. 
This work has been funded by the Deutsche Forschungsgemeinschaft (DFG, German Research Foundation) - 325594231, FOR 2634/2. T. H. and D.S. acknowledge support from the European Research Council under the
Horizon 2020 Framework Program via the ERC Advanced Grant Origins 83 24 28 (PI: Th. Henning). C.F.M. is funded by the European Union (ERC, WANDA, 101039452). Views and opinions expressed are however those of the author(s) only and do not necessarily reflect those of the European Union or the European Research Council Executive Agency. Neither the European Union nor the granting authority can be held responsible for them.
\end{acknowledgements}

\bibliography{biblio}

\begin{thebibliography}{31}
\expandafter\ifx\csname natexlab\endcsname\relax\def\natexlab#1{#1}\fi

\bibitem[{{Banzatti} {et~al.}(2022){Banzatti}, {Abernathy}, {Brittain},
  {Bosman}, {Pontoppidan}, {Boogert}, {Jensen}, {Carr}, {Najita}, {Grant},
  {Sigler}, {Sanchez}, {Kern}, \& {Rayner}}]{banzatti22a}
{Banzatti}, A., {Abernathy}, K.~M., {Brittain}, S., {et~al.} 2022, \aj, 163,
  174

\bibitem[{{Banzatti} {et~al.}(2012){Banzatti}, {Meyer}, {Bruderer}, {Geers},
  {Pascucci}, {Lahuis}, {Juh{\'a}sz}, {Henning}, \&
  {{\'A}brah{\'a}m}}]{banzatti12}
{Banzatti}, A., {Meyer}, M.~R., {Bruderer}, S., {et~al.} 2012, \apj, 745, 90

\bibitem[{{Banzatti} \& {Pontoppidan}(2015)}]{banzatti15}
{Banzatti}, A. \& {Pontoppidan}, K.~M. 2015, \apj, 809, 167

\bibitem[{{Banzatti} {et~al.}(2023){Banzatti}, {Pontoppidan}, {P{\'e}re
  Ch{\'a}vez}, {Salyk}, {Diehl}, {Bruderer}, {Herczeg}, {Carmona}, {Pascucci},
  {Brittain}, {Jensen}, {Grant}, {van Dishoeck}, {Kamp}, {Bosman}, {{\"O}berg},
  {Blake}, {Meyer}, {Gaidos}, {Boogert}, {Rayner}, \& {Wheeler}}]{banzatti23a}
{Banzatti}, A., {Pontoppidan}, K.~M., {P{\'e}re Ch{\'a}vez}, J., {et~al.} 2023,
  \aj, 165, 72

\bibitem[{{Banzatti} {et~al.}(2017){Banzatti}, {Pontoppidan}, {Salyk},
  {Herczeg}, {van Dishoeck}, \& {Blake}}]{banzatti17}
{Banzatti}, A., {Pontoppidan}, K.~M., {Salyk}, C., {et~al.} 2017, \apj, 834,
  152

\bibitem[{{Bast} {et~al.}(2011){Bast}, {Brown}, {Herczeg}, {van Dishoeck}, \&
  {Pontoppidan}}]{bast11}
{Bast}, J.~E., {Brown}, J.~M., {Herczeg}, G.~J., {van Dishoeck}, E.~F., \&
  {Pontoppidan}, K.~M. 2011, \aap, 527, A119

\bibitem[{{Blake} \& {Boogert}(2004)}]{blake04}
{Blake}, G.~A. \& {Boogert}, A.~C.~A. 2004, \apjl, 606, L73

\bibitem[{{Brittain} {et~al.}(2016){Brittain}, {Najita}, {Carr},
  {{\'A}d{\'a}mkovics}, \& {Reynolds}}]{brittain16}
{Brittain}, S.~D., {Najita}, J.~R., {Carr}, J.~S., {{\'A}d{\'a}mkovics}, M., \&
  {Reynolds}, N. 2016, \apj, 830, 112

\bibitem[{{Brittain} {et~al.}(2007){Brittain}, {Simon}, {Najita}, \&
  {Rettig}}]{brittain07a}
{Brittain}, S.~D., {Simon}, T., {Najita}, J.~R., \& {Rettig}, T.~W. 2007, \apj,
  659, 685

\bibitem[{{Brown} {et~al.}(2013){Brown}, {Pontoppidan}, {van Dishoeck},
  {Herczeg}, {Blake}, \& {Smette}}]{brown13}
{Brown}, J.~M., {Pontoppidan}, K.~M., {van Dishoeck}, E.~F., {et~al.} 2013,
  \apj, 770, 94

\bibitem[{{Dorn} {et~al.}(2023){Dorn}, {Bristow}, {Smoker}, {Rodler}, {Lavail},
  {Accardo}, {van den Ancker}, {Baade}, {Baruffolo}, {Courtney-Barrer},
  {Blanco}, {Brucalassi}, {Cumani}, {Follert}, {Haimerl}, {Hatzes}, {Haug},
  {Heiter}, {Hinterschuster}, {Hubin}, {Ives}, {Jung}, {Jones}, {Kaeufl},
  {Kirchbauer}, {Klein}, {Kochukhov}, {Korhonen}, {K{\"o}hler}, {Lizon},
  {Moins}, {Molina-Conde}, {Marquart}, {Neeser}, {Oliva}, {Pallanca},
  {Pasquini}, {Paufique}, {Piskunov}, {Reiners}, {Schneller}, {Schmutzer},
  {Seemann}, {Slumstrup}, {Smette}, {Stegmeier}, {Stempels}, {Tordo},
  {Valenti}, {Valenzuela}, {Vernet}, {Vinther}, \& {Wehrhahn}}]{dorn23}
{Dorn}, R.~J., {Bristow}, P., {Smoker}, J.~V., {et~al.} 2023, \aap, 671, A24

\bibitem[{{Fedele} {et~al.}(2011){Fedele}, {Pascucci}, {Brittain}, {Kamp},
  {Woitke}, {Williams}, {Dent}, \& {Thi}}]{fedele11}
{Fedele}, D., {Pascucci}, I., {Brittain}, S., {et~al.} 2011, \apj, 732, 106

\bibitem[{{Freudling} {et~al.}(2013){Freudling}, {Romaniello}, {Bramich},
  {Ballester}, {Forchi}, {Garc{\'{\i}}a-Dabl{\'o}}, {Moehler}, \&
  {Neeser}}]{reflex}
{Freudling}, W., {Romaniello}, M., {Bramich}, D.~M., {et~al.} 2013, \aap, 559,
  A96

\bibitem[{{Goldsmith} \& {Langer}(1999)}]{goldsmith99}
{Goldsmith}, P.~F. \& {Langer}, W.~D. 1999, \apj, 517, 209

\bibitem[{{Gravity Collaboration} {et~al.}(2021){Gravity Collaboration},
  {Perraut}, {Labadie}, {Bouvier}, {M{\'e}nard}, {Klarmann}, {Dougados},
  {Benisty}, {Berger}, {Bouarour}, {Brandner}, {Caratti O Garatti}, {Caselli},
  {de Zeeuw}, {Garcia-Lopez}, {Henning}, {Sanchez-Bermudez}, {Sousa}, {van
  Dishoeck}, {Al{\'e}cian}, {Amorim}, {Cl{\'e}net}, {Davies}, {Drescher},
  {Duvert}, {Eckart}, {Eisenhauer}, {F{\"o}rster-Schreiber}, {Garcia},
  {Gendron}, {Genzel}, {Gillessen}, {Grellmann}, {Hei{\ss}el}, {Hippler},
  {Horrobin}, {Hubert}, {Jocou}, {Kervella}, {Lacour}, {Lapeyr{\`e}re}, {Le
  Bouquin}, {L{\'e}na}, {Lutz}, {Ott}, {Paumard}, {Perrin}, {Scheithauer},
  {Shangguan}, {Shimizu}, {Stadler}, {Straub}, {Straubmeier}, {Sturm},
  {Tacconi}, {Vincent}, {von Fellenberg}, \& {Widmann}}]{gravity21}
{Gravity Collaboration}, {Perraut}, K., {Labadie}, L., {et~al.} 2021, \aap,
  655, A73

\bibitem[{{Gupta} {et~al.}(2023){Gupta}, {Miotello}, {Manara}, {Williams},
  {Facchini}, {Beccari}, {Birnstiel}, {Ginski}, {Hacar}, {K{\"u}ffmeier},
  {Testi}, {Tychoniec}, \& {Yen}}]{gupta23}
{Gupta}, A., {Miotello}, A., {Manara}, C.~F., {et~al.} 2023, \aap, 670, L8

\bibitem[{{Herczeg} {et~al.}(2011){Herczeg}, {Brown}, {van Dishoeck}, \&
  {Pontoppidan}}]{herczeg11}
{Herczeg}, G.~J., {Brown}, J.~M., {van Dishoeck}, E.~F., \& {Pontoppidan},
  K.~M. 2011, \aap, 533, A112

\bibitem[{{Kausch} {et~al.}(2015){Kausch}, {Noll}, {Smette}, {Kimeswenger},
  {Barden}, {Szyszka}, {Jones}, {Sana}, {Horst}, \& {Kerber}}]{kausch15}
{Kausch}, W., {Noll}, S., {Smette}, A., {et~al.} 2015, \aap, 576, A78

\bibitem[{{Komarova} \& {Fischer}(2020)}]{komarova20}
{Komarova}, O. \& {Fischer}, W.~J. 2020, Research Notes of the American
  Astronomical Society, 4, 6

\bibitem[{{Mandell} {et~al.}(2012){Mandell}, {Bast}, {van Dishoeck}, {Blake},
  {Salyk}, {Mumma}, \& {Villanueva}}]{mandell12}
{Mandell}, A.~M., {Bast}, J., {van Dishoeck}, E.~F., {et~al.} 2012, \apj, 747,
  92

\bibitem[{{Martin} {et~al.}(2018){Martin}, {Fitzgerald}, {McLean}, {Doppmann},
  {Kassis}, {Aliado}, {Canfield}, {Johnson}, {Kress}, {Lanclos}, {Magnone},
  {Sohn}, {Wang}, \& {Weiss}}]{martin18}
{Martin}, E.~C., {Fitzgerald}, M.~P., {McLean}, I.~S., {et~al.} 2018, in
  Society of Photo-Optical Instrumentation Engineers (SPIE) Conference Series,
  Vol. 10702, Ground-based and Airborne Instrumentation for Astronomy VII, ed.
  C.~J. {Evans}, L.~{Simard}, \& H.~{Takami}, 107020A

\bibitem[{{Najita} {et~al.}(2003){Najita}, {Carr}, \& {Mathieu}}]{najita03}
{Najita}, J., {Carr}, J.~S., \& {Mathieu}, R.~D. 2003, \apj, 589, 931

\bibitem[{{Rayner} {et~al.}(2022){Rayner}, {Tokunaga}, {Jaffe}, {Bond},
  {Bonnet}, {Ching}, {Connelley}, {Cushing}, {Kokubun}, {Lockhart}, {Vacca}, \&
  {Warmbier}}]{rayner22}
{Rayner}, J., {Tokunaga}, A., {Jaffe}, D., {et~al.} 2022, \pasp, 134, 015002

\bibitem[{{Rayner} {et~al.}(2016){Rayner}, {Tokunaga}, {Jaffe}, {Bonnet},
  {Ching}, {Connelley}, {Kokubun}, {Lockhart}, \& {Warmbier}}]{rayner16}
{Rayner}, J., {Tokunaga}, A., {Jaffe}, D., {et~al.} 2016, in Society of
  Photo-Optical Instrumentation Engineers (SPIE) Conference Series, Vol. 9908,
  Ground-based and Airborne Instrumentation for Astronomy VI, ed. C.~J.
  {Evans}, L.~{Simard}, \& H.~{Takami}, 990884

\bibitem[{{Rigliaco} {et~al.}(2015){Rigliaco}, {Pascucci}, {Duchene},
  {Edwards}, {Ardila}, {Grady}, {Mendigut{\'\i}a}, {Montesinos}, {Mulders},
  {Najita}, {Carpenter}, {Furlan}, {Gorti}, {Meijerink}, \&
  {Meyer}}]{rigliaco15}
{Rigliaco}, E., {Pascucci}, I., {Duchene}, G., {et~al.} 2015, \apj, 801, 31

\bibitem[{{Salyk} {et~al.}(2013){Salyk}, {Herczeg}, {Brown}, {Blake},
  {Pontoppidan}, \& {van Dishoeck}}]{salyk13}
{Salyk}, C., {Herczeg}, G.~J., {Brown}, J.~M., {et~al.} 2013, \apj, 769, 21

\bibitem[{{Smette} {et~al.}(2015){Smette}, {Sana}, {Noll}, {Horst}, {Kausch},
  {Kimeswenger}, {Barden}, {Szyszka}, {Jones}, {Gallenne}, {Vinther},
  {Ballester}, \& {Taylor}}]{smette15}
{Smette}, A., {Sana}, H., {Noll}, S., {et~al.} 2015, \aap, 576, A77

\bibitem[{{Sullivan} {et~al.}(2019){Sullivan}, {Prato}, {Edwards}, {Avilez}, \&
  {Schaefer}}]{sullivan19}
{Sullivan}, K., {Prato}, L., {Edwards}, S., {Avilez}, I., \& {Schaefer}, G.~H.
  2019, \apj, 884, 28

\bibitem[{{Thi} {et~al.}(2013){Thi}, {Kamp}, {Woitke}, {van der Plas},
  {Bertelsen}, \& {Wiesenfeld}}]{thi13}
{Thi}, W.~F., {Kamp}, I., {Woitke}, P., {et~al.} 2013, \aap, 551, A49

\bibitem[{{Ulmer-Moll} {et~al.}(2019){Ulmer-Moll}, {Figueira}, {Neal},
  {Santos}, \& {Bonnefoy}}]{ulmermoll19}
{Ulmer-Moll}, S., {Figueira}, P., {Neal}, J.~J., {Santos}, N.~C., \&
  {Bonnefoy}, M. 2019, \aap, 621, A79

\bibitem[{{van der Plas} {et~al.}(2015){van der Plas}, {van den Ancker},
  {Waters}, \& {Dominik}}]{vanderplas11}
{van der Plas}, G., {van den Ancker}, M.~E., {Waters}, L.~B.~F.~M., \&
  {Dominik}, C. 2015, \aap, 574, A75

\end{thebibliography}

\newpage

\appendix

\section{Observing details}
The observing conditions for our observations are given in Table~\ref{tab: obs log}. 

\begin{table*}[]
    \centering
    \begin{tabular}{c|cccc|cccc}
    \hline\hline
    & \multicolumn{4}{c|}{S CrA} & \multicolumn{4}{c}{$\lambda$ Aql} \\\cline{1-9}  
    Filter & Time & Seeing & Airmass & I.Q. & Time & Seeing & Airmass & I.Q. \\
    & [UT] & [$^{\prime\prime}$] & & [$^{\prime\prime}$] & [UT] & [$^{\prime\prime}$] & & [$^{\prime\prime}$] \\\cline{0-8}  
L3244 & 2021-09-18T02:51:58.8391 & 0.7 & 1.295 & 1.0 & 2021-09-18T03:51:09.7134 & 1.11 & 1.926 & 0.84\\ 
L3262 & 2021-09-18T03:00:15.5436 & 0.94 & 1.327 & 1.22 & 2021-09-18T03:56:48.2035 & 0.98 & 1.999 & 0.85\\ 
L3302 & 2021-09-18T03:07:22.7153 & 1.05 & 1.357 & 1.03 & 2021-09-18T04:21:26.7844 & 1.31 & 2.413 & 0.93\\ 
L3340, P.A. 1 & 2021-09-18T02:37:17.0365 & 0.68 & 1.245 & 0.91 & 2021-09-18T03:45:20.0723 & 1.22 & 1.857 & 1.02\\ 
L3340, P.A. 2 & 2021-09-18T02:44:34.2579 & 0.63 & 1.269 & 0.93 & 2021-09-18T03:45:20.0723 & 1.22 & 1.857 & 1.02\\ 
L3377 & 2021-09-18T03:15:12.8397 & 0.7 & 1.392 & 0.94 & 2021-09-18T04:27:17.7477 & 1.02 & 2.542 & 0.72\\ 
L3412 & 2021-09-18T03:22:25.3378 & 0.91 & 1.428 & 1.08 & 2021-09-18T04:32:54.9032 & 1.05 & 2.682 & 0.73\\ 
L3426 & 2021-09-18T03:29:31.5921 & 0.86 & 1.465 & 0.96 & 2021-09-18T04:38:11.7044 & 1.06 & 2.828 & 0.83\\ 
M4187 & 2021-09-18T23:44:52.1024 & 0.43 & 1.024 & 0.77 & 2021-09-19T01:08:23.3365 & 0.29 & 1.115 & 0.63\\ 
M4211 & 2021-09-18T23:52:33.3342 & 0.43 & 1.023 & 0.77 & 2021-09-19T01:16:29.1057 & 0.33 & 1.128 & 0.65\\ 
M4266 & 2021-09-19T00:00:24.7822 & 0.31 & 1.024 & 0.75 & 2021-09-19T01:26:28.1670 & 0.39 & 1.147 & 0.67\\ 
M4318 & 2021-09-19T00:08:30.8910 & 0.29 & 1.025 & 0.73 & 2021-09-19T01:34:27.2257 & 0.25 & 1.163 & 0.67\\ 
M4368, P.A. 1 & 2021-09-18T23:16:27.7131 & 0.51 & 1.033 & 0.73 & 2021-09-19T01:00:22.7853 & 0.33 & 1.104 & 0.68\\ 
M4368, P.A. 2 & 2021-09-18T23:36:45.9149 & 0.36 & 1.025 & 0.76 & 2021-09-19T01:00:22.7853 & 0.33 & 1.104 & 0.68\\ 
M4416 & 2021-09-19T00:18:05.5455 & 0.32 & 1.028 & 0.77 & 2021-09-19T01:43:49.4897 & 0.34 & 1.186 & 0.61\\ 
M4461 & 2021-09-19T00:26:07.9014 & 0.49 & 1.032 & 0.73 & 2021-09-19T01:51:10.0201 & 0.33 & 1.205 & 0.66\\ 
M4504 & 2021-09-19T00:34:53.1912 & 0.33 & 1.037 & 0.73 & 2021-09-19T01:58:46.7159 & 0.28 & 1.227 & 0.59\\ 
M4519 & 2021-09-19T00:42:57.2216 & 0.28 & 1.043 & 0.74 & 2021-09-19T02:06:01.9413 & 0.29 & 1.25 & 0.6\\ 
    \hline \hline
    \end{tabular}
    \caption{Observing time, seeing, airmass, and airmass corrected seeing (I.Q.) for our observations. }
    \label{tab: obs log}
\end{table*}

\section{Full spectra of S CrA N and S}
\renewcommand\thefigure{\thesection.\arabic{figure}}  
\setcounter{figure}{0} 
The full L- and M-band spectra for S CrA N are shown in Figure~\ref{fig: overview Lband} and Figure~\ref{fig: overview Mband}, respectively. And the S CrA S L- and M-band spectra are presented in Figures~\ref{fig: overview Lband S CrA S} and \ref{fig: overview Mband S CrA S}, respectively. The oCRIRES data for each binary component are also shown for comparison. 

\begin{figure*}[]
\includegraphics[scale=0.51]{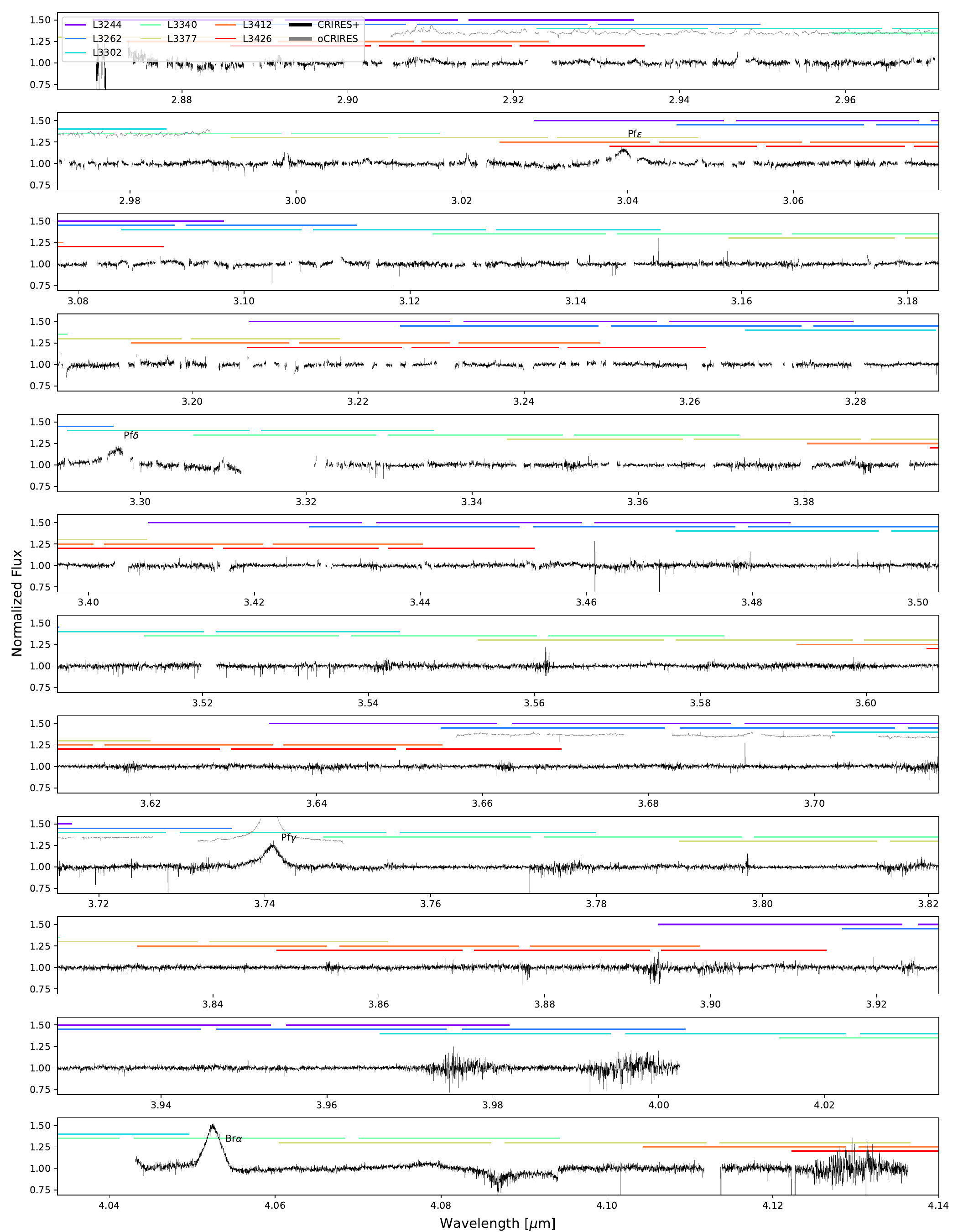}
\caption{S CrA N spectra after combining all of the L-band filters (black). For comparison, the oCRIRES data, taken with a longer total exposure time, are shown in grey, vertically offset for clarity. Wavelength regions where the telluric lines are particularly strong have been removed. The wavelength coverage of the L-band spectral settings is shown in the colors. Detected hydrogen recombination lines are labeled. 
     \label{fig: overview Lband}}
\end{figure*}

\begin{figure*}[]
\includegraphics[scale=0.51]{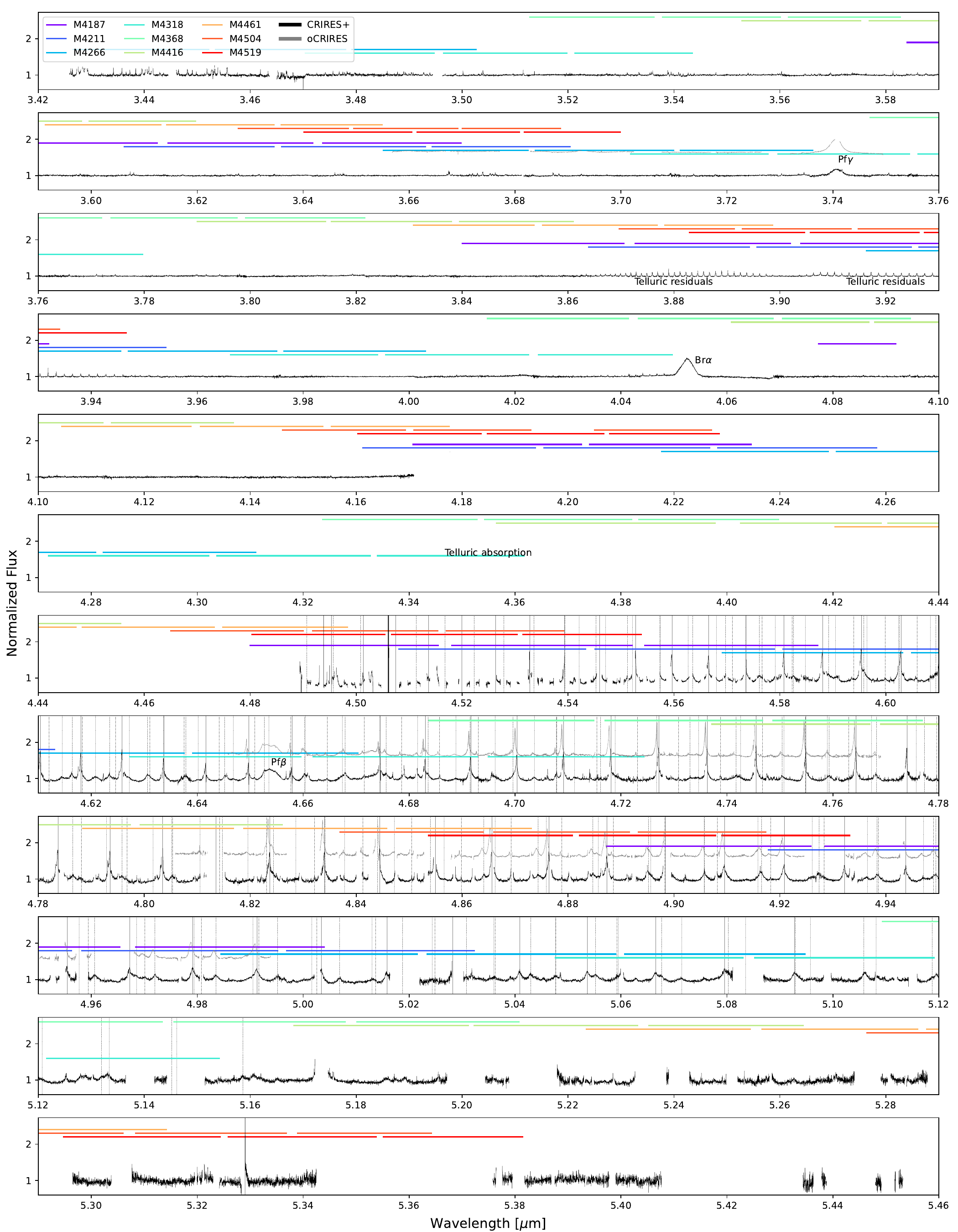}
\caption{The same as Figure~\ref{fig: overview Lband}, but showing the combination of all of the M-band filters. Vertical grey lines indicate CO ro-vibrational lines: $^{12}$CO v=1-0 (solid), v=2-1 (dashed), v=3-2 (dotted), and $^{13}$CO v=1-0 (dot-dashed). 
     \label{fig: overview Mband}}
\end{figure*}

\begin{figure*}[]
\includegraphics[scale=0.51]{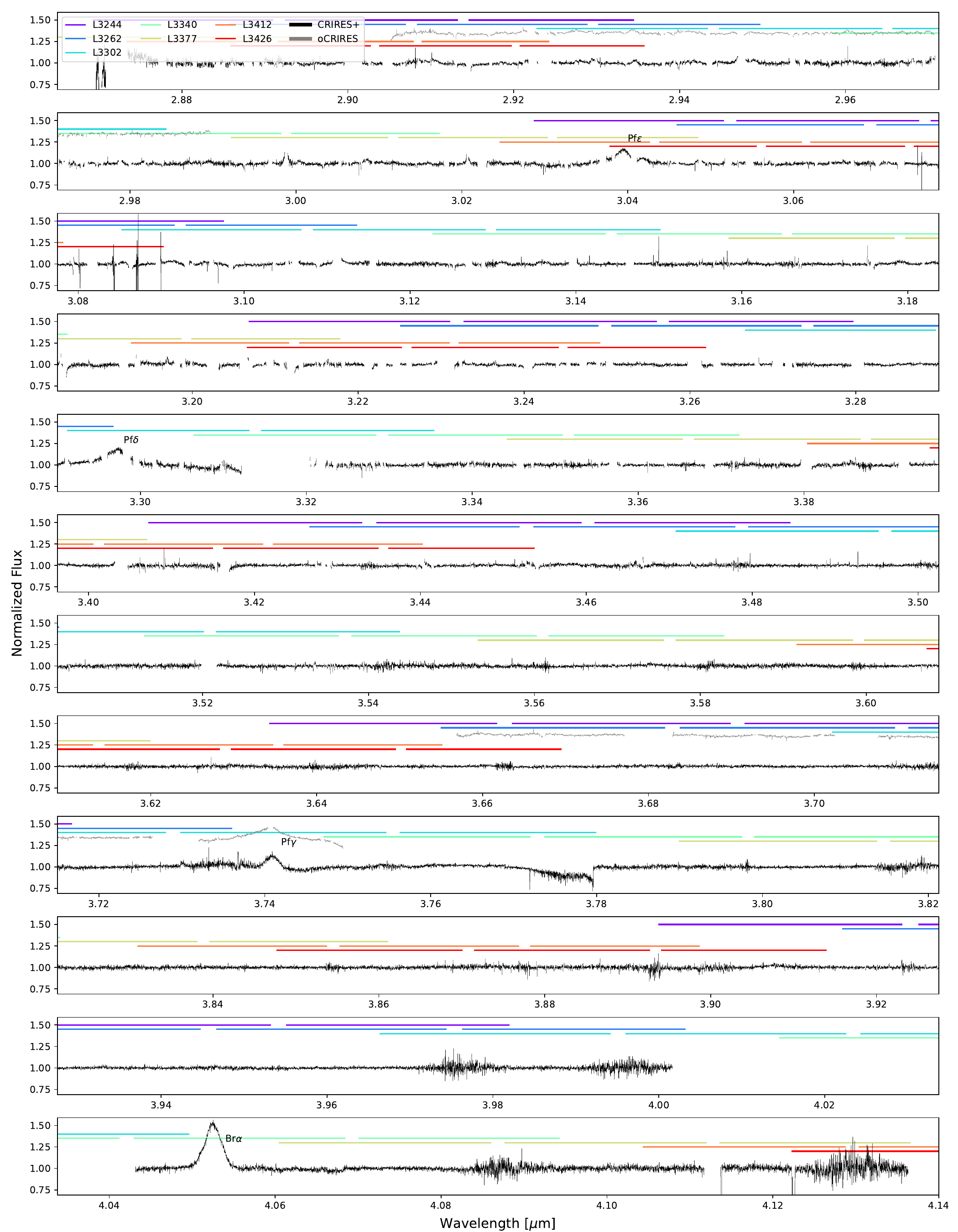}
\caption{L-band data for S CrA S.
     \label{fig: overview Lband S CrA S}}
\end{figure*}

\begin{figure*}[]
\includegraphics[scale=0.51]{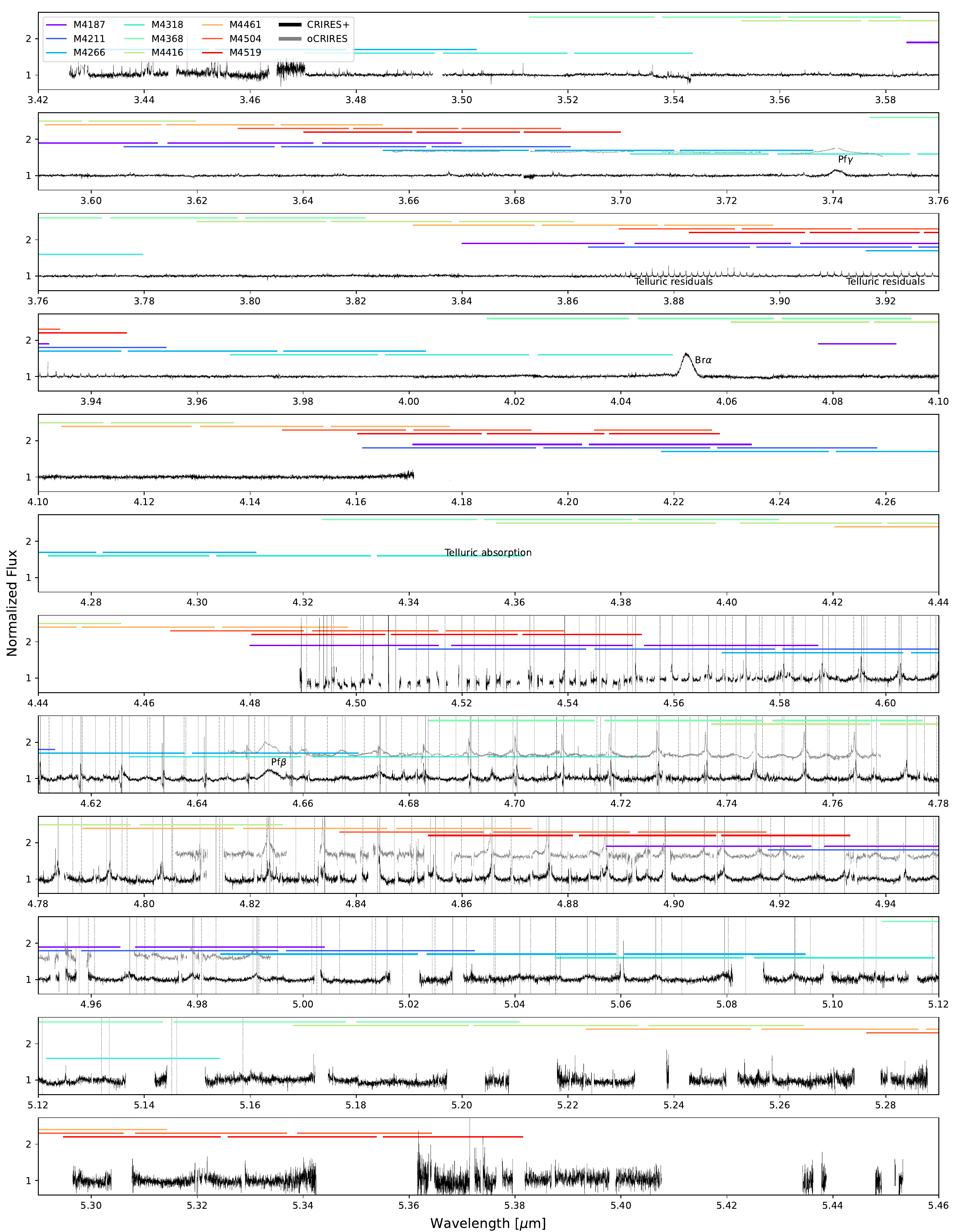}
\caption{M-band data for S CrA S.
     \label{fig: overview Mband S CrA S}}
\end{figure*}

\end{document}